\documentclass[preprint]{iucr}              %

     \journalcode{A}              %

\usepackage{gensymb}
\usepackage{url}
\usepackage{enumitem}
\usepackage{xspace}
\usepackage{fancyhdr}
\usepackage{afterpage}
\usepackage{ulem}
\usepackage[usenames,dvipsnames]{xcolor}
\usepackage{dcolumn}
\usepackage{comment}
\usepackage{listings}
\usepackage{makeidx}
\usepackage{longtable}
\usepackage[utf8]{inputenc}
\usepackage{comment}

\usepackage[dvips]{epsfig}
\usepackage{epstopdf}
\usepackage{graphicx}%

\usepackage{booktabs}

\usepackage{mhchem}

\usepackage{mathtools}
\usepackage{amsmath}
\usepackage{amssymb}
\usepackage{bm}
\usepackage{amsfonts}
\usepackage{commath} %
\usepackage[a4paper, total={6in, 9in}]{geometry}

\usepackage{subfiles}

\newcommand{\sjba}[1]{}

\newcommand{\mta}[1]{}

\newcommand{\eq}[1]{Eq.~\ref{eq:#1}}

\newcommand{\fig}[1]{Fig.~\ref{fig:#1}}

\newcommand{\sect}[1]{Section~\ref{sec:#1}}

\graphicspath{ {./figures/} }

\definecolor{dgreen}{HTML}{008000}

\definecolor{codegreen}{rgb}{0,0.6,0}
\definecolor{codegray}{rgb}{0.5,0.5,0.5}
\definecolor{codepurple}{rgb}{0.58,0,0.82}
\definecolor{backcolour}{rgb}{0.95,0.95,0.92}
\lstdefinestyle{mystyle}{
    backgroundcolor=\color{backcolour},
    commentstyle=\color{codegreen},
    keywordstyle=\color{magenta},
    numberstyle=\tiny\color{codegray},
    stringstyle=\color{codepurple},
    basicstyle=\ttfamily\footnotesize,
    breakatwhitespace=false,
    breaklines=true,
    captionpos=b,
    keepspaces=true,
    numbers=left,
    numbersep=5pt,
    showspaces=false,
    showstringspaces=false,
    showtabs=false,
    tabsize=2
}
\newcounter{saveenumi}

\newcommand{\be}{\begin{enumerate}}
\newcommand{\ee}{\end{enumerate}}
\newcommand{\bes}{\begin{enumerate}[wide, labelwidth=!, labelindent=0pt, label=\textbf{\textcolor{blue}{\arabic*}.}]}
\newcommand{\ees}{\end{enumerate}}

\usepackage{float}
\newcommand{\vphi}{\ensuremath{\varphi_0}\xspace}
\newcommand{\phimax}{\ensuremath{\vphi^*}\xspace}
\newcommand{\phimaxone}{\ensuremath{\vphi^{*,1}}\xspace}
\newcommand{\phimaxtwo}{\ensuremath{\vphi^{*,2}}\xspace}
\newcommand{\fwhm}{\ensuremath{\Delta\varphi}\xspace}

\newcommand{\skl}{SkL\xspace}
\newcommand{\nfolds}{\ensuremath{n_{\text{folds}}}\xspace}
\newcommand{\ddoverlap}{\ensuremath{\frac{d^2\mathcal{O}}{d\vphi^2}}\xspace}
\newcommand{\intensity}{\ensuremath{\mathbf{I}}\xspace}
\newcommand{\mask}{\textbf{M}\xspace}
\newcommand{\mapmask}{\ensuremath{\mathbf{M}_{map}}\xspace}
\newcommand{\atomap}{{\sc Atomap}\xspace}
\newcommand{\threshold}{\ensuremath{\tau}\xspace}

\newcommand{\overlap}{\ensuremath{\mathcal{O}(\vphi)}\xspace}
\newcommand{\overlapsub}[1]{\ensuremath{\mathcal{O}_{#1}(\vphi)}\xspace}
\newcommand{\overlapeq}{\ensuremath{\mathcal{O}(\vphi) = \SAF\cdot\intensity}}

\newcommand{\SAF}{\ensuremath{\mathbf{S}({\vphi})}\xspace}
\newcommand{\SDF}{\ensuremath{\textbf{R}(r, \varphi; \vphi)}\xspace}
\newcommand{\Sof}[1]{\ensuremath{\mathbf{S}({#1})}\xspace}

\begin{document}                  %

\title{High-Throughput Orientation Extraction via Symmetry-Adapted Filters}


\author[a]{Caden}{Myers}{}{}
\author[a]{Yucong}{Chen}{}{}
\author[b]{Nathan}{Chalus}{}{}
\author[c]{Tess}{Smidt}{}{}
\author[b]{Morten~R.}{Eskildsen}{}{}
\cauthor[a,d]{Simon~J.~L.}{Billinge}{sbillinge@ucsb.edu}{}

\aff[a]{Department of Applied Physics \& Applied Mathematics, Columbia University, \city{New York}, NY~10025, \country{USA}}
\aff[b]{Department of Physics \& Astronomy, University of Notre Dame, \city{Notre Dame}, IN~46556, \country{USA}}
\aff[c]{Department of Electrical Engineering and Computer Science, Massachusetts Institute of Technology, \city{Cambridge}, MA~02139, \country{USA}}
\aff[d]{Department of Materials, University of California, Santa Barbara, \city{Santa Barbara}, CA~93106, \country{USA}}
\shortauthor{Myers et al.}

\maketitle                        %

\begin{abstract}

An algorithm for finding the orientation of objects in an image that have a known rotational symmetry, the symmetry-adapted filter (SAF) algorithm, is presented.
It is deterministic and training-free, which makes it fast and reproducible.
A filter that encodes the known rotational symmetry and approximate peak width is centered on the symmetry center of the object of interest, masked to the region of interest, and rotated to maximize its inner product with the image.
The approach is demonstrated on more than 20,000 small-angle neutron scattering images of a torque-driven skyrmion lattice.
It is also demonstrated on atomically resolved real-space images, where mapping local atomic orientations enabled the identification of grains and grain boundaries and the quantification of lattice relaxation.

\end{abstract}

\section{Introduction}

Experiments that use area detectors can rapidly generate large two-dimensional datasets, particularly in X-ray and neutron scattering experiments at large-scale facilities and in electron imaging and diffraction techniques~\cite{forsterTransformingXrayDetection2019, wangSynchrotronBigData2018, muhlbauerKineticSmallAngle2016, spurgeonDataDrivenNextgenerationTransmission2021, ophusFourDimensionalScanning2019}.
For ordered samples, these datasets often show the rotational symmetry of the underlying structure~\cite{alsnielsenElementsModernXray2011}.
In certain applications, the orientation angle of these symmetric features is itself the quantity of interest.
For example, skyrmions in chiral magnets such as MnSi and \ce{Cu2OSeO3} can order into a hexagonal skyrmion lattice (SkL)~\cite{muhlbauerSkyrmionLatticeChiral2009, sekiObservationSkyrmionsMultiferroic2012, nagaosaTopologicalPropertiesDynamics2013}.
In small-angle neutron scattering (SANS), the hexagonal SkL appears as six Bragg peaks on a ring on an area detector~\cite{muhlbauerSkyrmionLatticeChiral2009, adamsLongWavelengthHelimagneticOrder2012, MuhlbauerMagneticSANS2019}.
The angular position of these peaks gives a direct measure of the in-plane orientation of the SkL.
The SkL can be rotated by electric currents combined with thermal gradients~\cite{jonietzSpinTransferTorques2010, everschorRotatingSkyrmionLattices2012, chalusSkyrmionLatticeManipulation2025}, by thermal gradients alone~\cite{mochizukiThermallyDrivenRatchet2014, pollathDynamicalDefectsRotating2017}, and by electric fields~\cite{whiteElectricFieldInducedSkyrmion2014}.
Tracking the orientation of the Bragg peaks over time with SANS therefore reveals the rotational dynamics of the SkL~\cite{muhlbauerKineticSmallAngle2016, whiteElectricFieldInducedSkyrmion2014, chalusSkyrmionLatticeManipulation2025}.

Another case where orientation angle is important is in real-space imaging.
In atomically resolved transmission electron microscope (TEM) images, the local orientation of the lattice around each atom reveals grains, grain boundaries, and strainwithin the lattice~\cite{hytchQuantitativeMeasurementDisplacement1998, galindoPeakPairsAlgorithm2007, berkelsIdentificationGrainBoundary2007}.
Here, the symmetric feature is the ring of nearest-neighbor atoms around each atom.
In this case, the orientation angle of the local atomic neighborhood is measured at every atomic site rather than once per image.
Mapping these local orientations is commonly done with geometric phase analysis, the peak pairs method, or the bond-orientational order parameter computed from the atom positions~\cite{hytchQuantitativeMeasurementDisplacement1998, galindoPeakPairsAlgorithm2007, nelsonDislocationMediatedMelting1979}.

In both SANS and TEM, data are often imperfect, making extraction of orientation angles challenging.
Sample disorder and experimental effects can make certain peaks indeterminate and introduce noise to the data.
In SANS patterns, the signals can be weak or smeared depending on sample alignment and disorder~\cite{muhlbauerSkyrmionLatticeChiral2009, MuhlbauerMagneticSANS2019}.
Additionally, several domains with different orientations can show up in the same pattern~\cite{bannenbergReorientationsRelaxationsMetastabilities2017, pollathDynamicalDefectsRotating2017}.
In TEM images, the atomic columns can be noisy or blurred.
Low electron doses, used to limit beam damage, leave the images noisy~\cite{egertonRadiationDamageOrganic2019}, and sample drift and scan distortions can shift and stretch the lattice~\cite{ophusCorrectingNonlinearDrift2016}.
Near grain boundaries and defects, the ring of neighboring atoms is also distorted or contains atoms from two grains, so the local symmetry is no longer perfect.
Manual analysis is manageable for small datasets, but with thousands of patterns or atomic sites, manual analysis becomes unrealistic and automated solutions to find orientation angles are needed.

Standard peak-finding and peak-fitting programs, such as {\sc SciPy}~\cite{scipy-reference} and GRASP~\cite{dewhurstGraphicalReductionAnalysis2023}, do a good job of finding peaks when they are well separated, but tend to fail when the peaks are noisy, smeared, or overlapping.
Peak fitting also needs starting positions set by hand, which is hard to scale to thousands of patterns.
Angular Fourier analysis of the intensity around the ring is another common model-free option~\cite{wochnerXrayCrossCorrelation2009}, but it faces the same challenges with noisy data and overlapping signals.

In this paper, we present a method for extracting orientation angles from noisy and incomplete data in 2D images by leveraging a known symmetry.
This approach, called the symmetry-adapted filter (SAF) algorithm, takes inspiration from the noise tolerance of machine learning models like convolutional neural networks and other image processing methods like template matching, circular harmonics, and steerable filters~\cite{brunelliTemplateMatchingTechniques2009, hsuRotationinvariantDigitalPattern1982, freemanDesignUseSteerable1991}.
The algorithm works by building a filter from a known or hypothesized rotational symmetry and approximate peak width, centering it, masking it to the region of interest, rotating it through a series of angles, and finding the angle which maximizes the inner product between the filter and the two-dimensional data.

We demonstrate the SAF algorithm by applying it to SANS data of a driven SkL in \ce{Cu2OSeO3}, extracting the angles of multiple noisy, symmetric signals from more than 20,000 SANS patterns~\cite{chalus2026therbino}.
We also applied the SAF algorithm to atomically resolved TEM
images to extract the orientation angle of the local atomic environment of every atom in the image, allowing for grain identification and strain effects to be observed.

\section{Motivation}
\label{sec:motivation}

The initial motivation for developing the algorithm discussed in this paper came from a study on skyrmion lattice (SkL) dynamics in \ce{Cu2OSeO3}.
In that study, a magnetic field and a radial thermal gradient were applied to the sample.
The radial thermal gradient causes the SkL to rotate~\cite{mochizukiThermallyDrivenRatchet2014, pollathDynamicalDefectsRotating2017, jinRollingMotionRigid2024, chalus2026therbino}.
By changing the magnetic field and the thermal gradient, the rotational dynamics of the SkLs change.

\begin{figure}\centering\includegraphics[width=\linewidth]{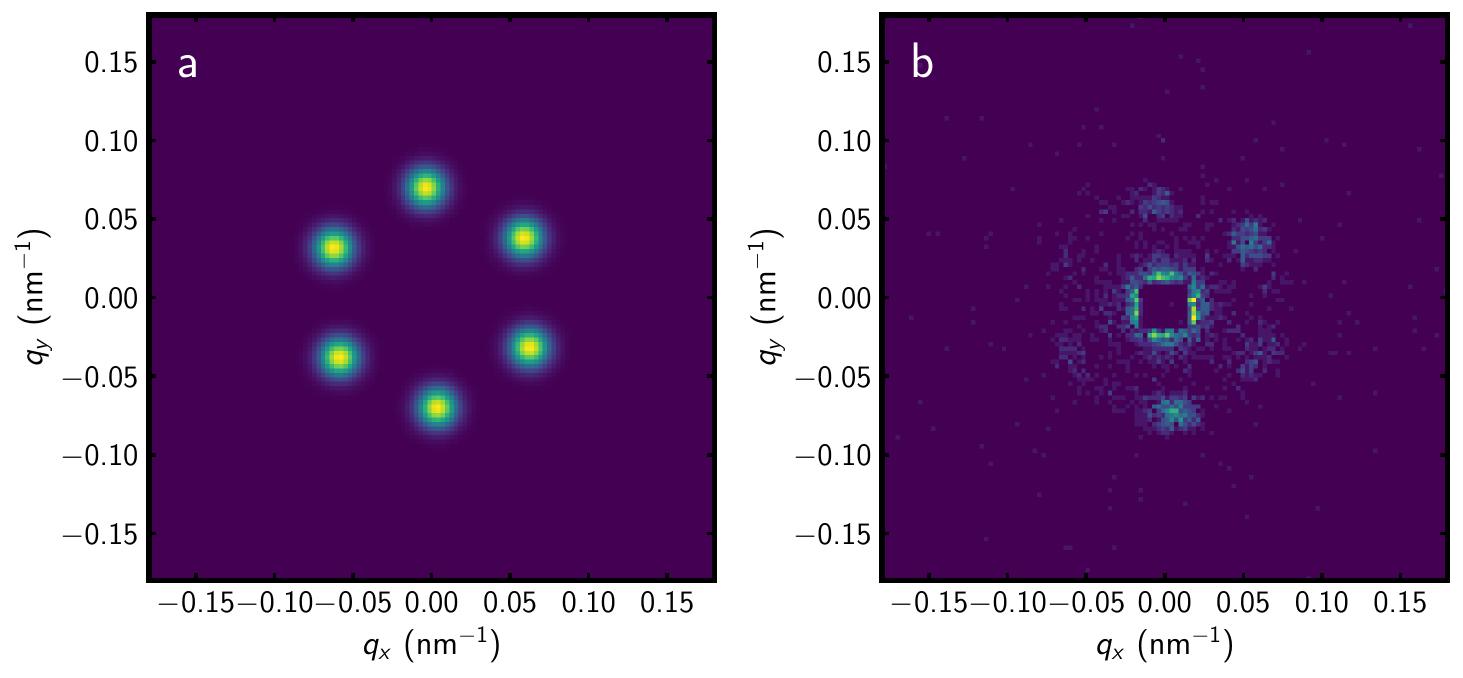}
    \caption{\textbf{(a)} The simulated SANS diffraction pattern of the  skyrmion lattice in our sample consisting of a sixfold symmetric arrangement of Gaussians. \textbf{(b)} A typical experimental SANS diffraction pattern from the skyrmion lattice under an applied thermal gradient.}
    \label{fig:ideal-vs-unideal-dps}
\end{figure}

To study the SkL rotation, small-angle neutron scattering (SANS) was used to collect time-series data.
The SANS measurement produced sets of 2D diffraction images.  The skyrmions form a triangular lattice.
Under steady-state and ideal conditions (static magnetic field, no radial thermal gradient, perfect SkL crystallinity, and perfect sample alignment), the SANS pattern exhibits clear sixfold rotational symmetry, as shown by the simulated pattern in \fig{ideal-vs-unideal-dps}(a).

However, in practice SANS measurements deviate from these ideal conditions.
Effects like thermal fluctuations, sample misalignment, reduced SkL crystallinity, weak scattering and other experimental imperfections degrade the resulting SANS pattern.
This results in noisy data containing broadened or weak/missing peaks. A more characteristic experimental pattern is shown in \fig{ideal-vs-unideal-dps}(b).

\begin{figure}
    \centering
\includegraphics[width=\linewidth]{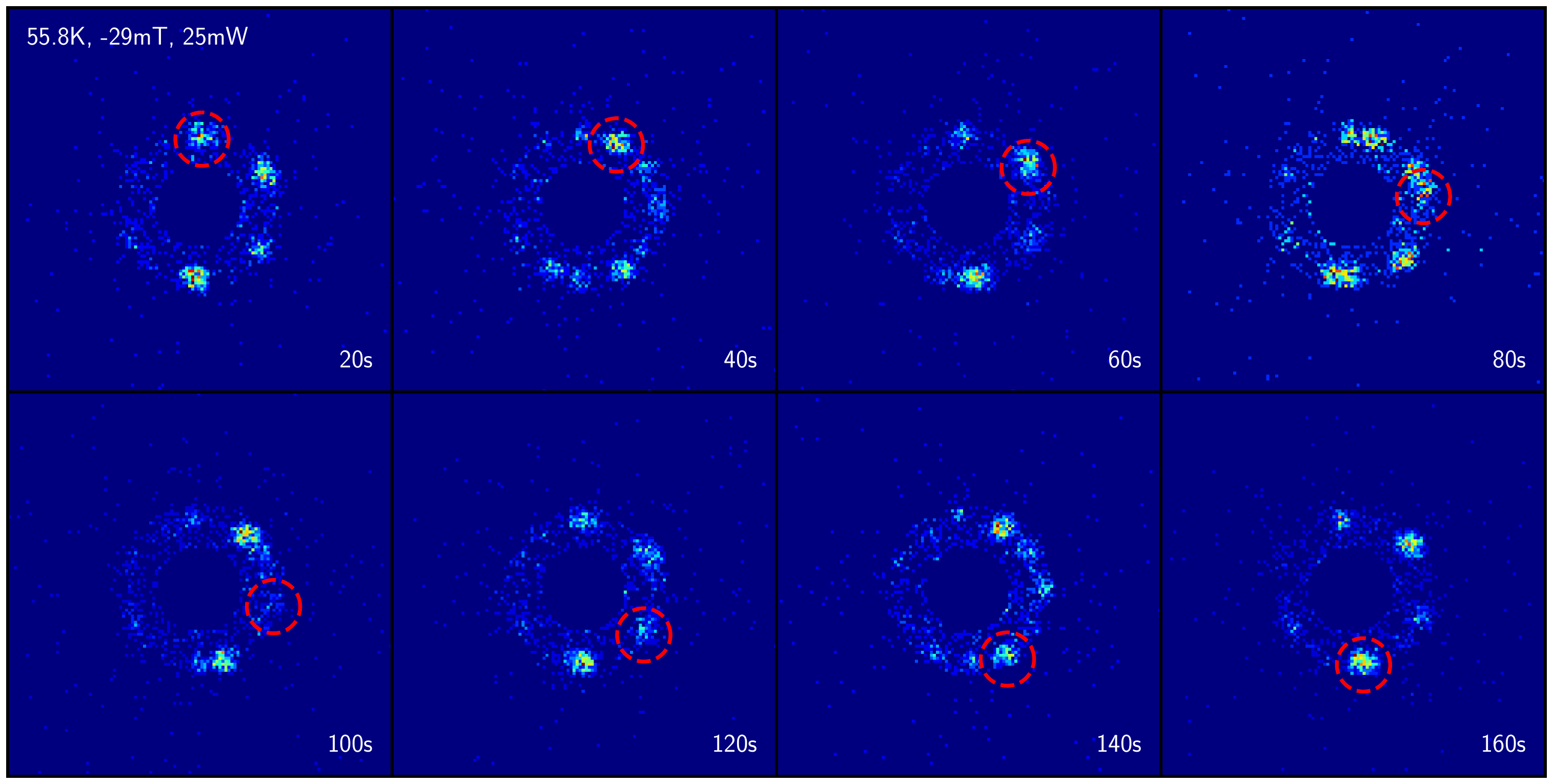}
    \caption{The time-series SANS patterns of the skyrmion lattices under a heating power of 25~mW. The dashed red circle is used to aid in visualizing the rotation in time of the strongest sixfold signal. In some frames, such as 40~s and 140~s, two sixfold signals can be seen.}
    \label{fig:rotating-skl}
\end{figure}

Additionally, when a radial thermal gradient is applied, the system is driven out of equilibrium causing the lattice, and therefore the sixfold signal in the SANS pattern, to rotate.
The SkL rotating under a radial thermal gradient is shown as a series of SANS patterns in \fig{rotating-skl}.
For this dataset, the driving force is clockwise, resulting in peaks that rotate in the same direction.
Since the peaks move steadily, a single peak can be tracked with the red dashed circle to see how it evolves in subsequent images.
As can be seen, the peak intensity and shape vary with position, which is one of the data interpretation challenges that we seek to solve.
Under driven conditions, peaks in the diffraction pattern may become smeared along the azimuthal direction.
We also found that the scattering pattern often displayed two coexisting sixfold signals with different orientation angles as shown in~\fig{rotating-skl}.
Therefore, to fully characterize the system dynamics, the objective became to extract the orientation angles of both sixfold diffraction signals as a function of time in an automated, model-independent manner.

The data present several challenges to achieve this objective.
First, the diffraction patterns are often noisy and imperfect.
Second, multiple sixfold signals with different orientation angles may be present simultaneously. And last, the datasets are too large for manual analysis.
These challenges prompted the development of the symmetry-adapted filter (SAF) algorithm described in this work.

The SAF algorithm was developed with inspiration from convolutional neural networks (CNNs), which are able to extract relevant information for a task from noisy image data.
The convolutional kernel is a mask that is multiplied onto patches of the image resulting in a product that is effective at accomplishing a classification task~\cite{Goodfellow2016deeplearning}.
CNN performance is robust against noise, small distortions, and variance in the image, recognizing features even when they are degraded.
We adopt a similar computational strategy to automate the extraction of SkL orientation from large time-series datasets, even when the diffraction pattern is degraded.
An important distinction to make between the SAF algorithm and CNNs is that the SAF algorithm does not involve any training or machine learning.
All steps in the SAF algorithm are fully deterministic and explicitly defined, guided by known physics of the problem.

The development and deployment of the SAF algorithm is described here.
The application of the SAF algorithm to study the dynamics of a driven SkL is presented in~\cite{chalus2026therbino}.

\section{Algorithm Overview}
\label{sec:algorithm-overview}

\begin{figure}
    \centering
    \includegraphics[width=\linewidth]{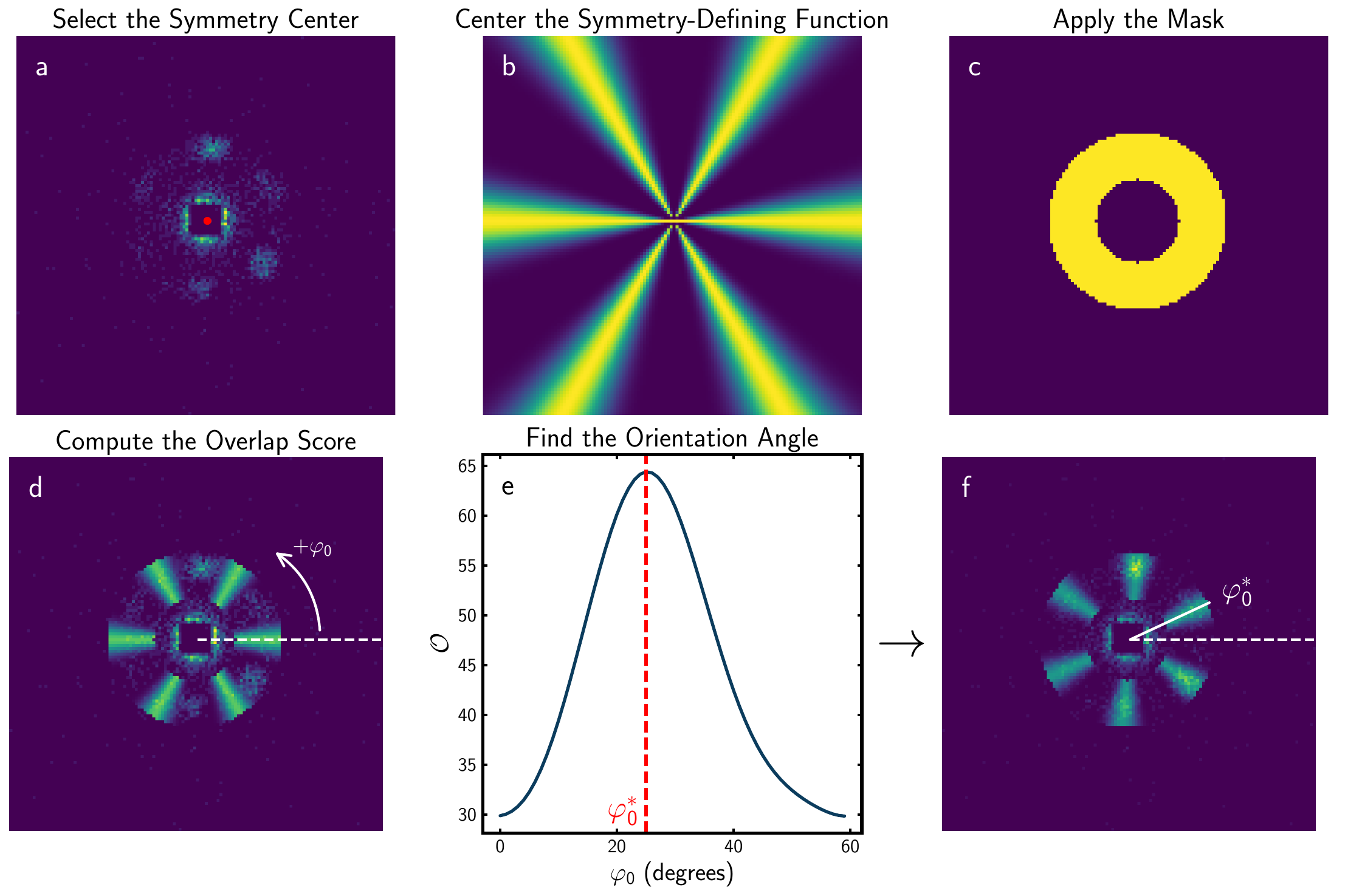}
    \caption{\textbf{(a)} A representative SANS pattern, \intensity, from the skyrmion lattice dataset. The red dot indicates the chosen center of rotational symmetry of the diffraction pattern.
    \textbf{(b)} The SDF computed on the same grid as \intensity, centered at the center of symmetry and evaluated at $\vphi = 0$.
    \textbf{(c)} The mask used to isolate the SANS signal.
    \textbf{(d)} The SAF evaluated at $\vphi = 0$, overlaid on the SANS pattern to visualize the computation of the overlap score.
    \textbf{(e)} The overlap score, \overlap, evaluated in the range $0^\circ < \vphi < 60^\circ$ in steps of $0.5^\circ$. The red dashed line marks the optimal orientation \phimax that maximizes $\mathcal{O}$.
    \textbf{(f)} The SAF evaluated at \phimax overlaid on the SANS pattern, illustrating how the SAF overlaps with the symmetric features in the data when evaluated at \phimax.
    }
    \label{fig:algorithm-overview}
\end{figure}

The objective of the SAF algorithm is to determine the orientation angle,~\phimax, of features with known symmetry in two-dimensional data, \intensity.
This is accomplished by computing a summed element-wise product of \intensity with one or more masks and a filter.
The product is evaluated as a function of the filter’s rotation angle.
We then assume that the angle that maximizes the product reveals the orientation of the symmetric features in \intensity.
To illustrate how the SAF algorithm accomplishes this task, we consider its application to a representative SANS pattern from the skyrmion lattice example in \fig{algorithm-overview}.
First, the center of rotational symmetry in \intensity is identified.
In \fig{algorithm-overview}(a), this center is indicated by a red dot near the middle of the image.
Then the Symmetry-Defining Function (SDF), denoted by \SDF, is computed on the same grid as \intensity for an initial phase angle, \vphi.
The SDF encodes the rotational symmetry of the features in~\intensity.
This is shown in \fig{algorithm-overview}(b) for $\vphi = 0$.
The SDF is described in more detail in \sect{sdf}.
Once the SDF has been computed, one or more masks, $\mask_i$, are applied to isolate regions of \intensity\ containing the signal of interest.
For each scientific use case, the masks may be designed for the particular physics of the situation.
The mask used for the SANS pattern is shown in \fig{algorithm-overview}(c).
Here, an annular mask is used with pixels inside the annulus being set to~1 and~0 elsewhere.
The mask isolates the radial range where the diffraction peaks are to be found in the SANS pattern.
Applying the mask to the SDF yields the SAF.
With the SAF defined, the inner product between the SAF and~\intensity is computed to obtain a scalar value as a function of~\vphi.
We call this function the overlap score,~\overlap.
The overlap score calculated for the given SANS pattern is shown in \fig{algorithm-overview}(e).
An overlay of the SAF rotated by~\phimax and the SANS pattern,~\intensity, is displayed in \fig{algorithm-overview}(f) showing how the SAF lines up with the Bragg peaks when $\vphi = \phimax$.

\section{Algorithm Details}
\label{sec:algorithm-details}

After introducing the main concept above we lay out the details of the algorithm.

\subsection{Symmetry-Defining Function (SDF) Design}
\label{sec:sdf}

When designing the SDF, we aim to encode known physics of the problem.  For example, in the \skl example, two key features are desired.
First, we want the SDF to encode the rotational symmetry and, second, we want it to encode information about the feature size of the diffraction peaks.
An SDF that has this desired behavior for the \skl case is \begin{equation}
    \SDF = \left|\cos\left(\frac{\nfolds}{2}(\varphi - \vphi)\right)\right|^k.\label{eq:sdf}
\end{equation}
In \eq{sdf}, \nfolds defines the degree of rotational symmetry (i.e.,~6 for the \skl),  \vphi defines the rotational offset about the defined center, and $k$ tunes the width of the peaks to match the size of the features in the data, as evident in \fig{algorithm-overview}(b), (d) and (f).

\begin{figure}
    \centering
    \includegraphics[width=0.7\linewidth]{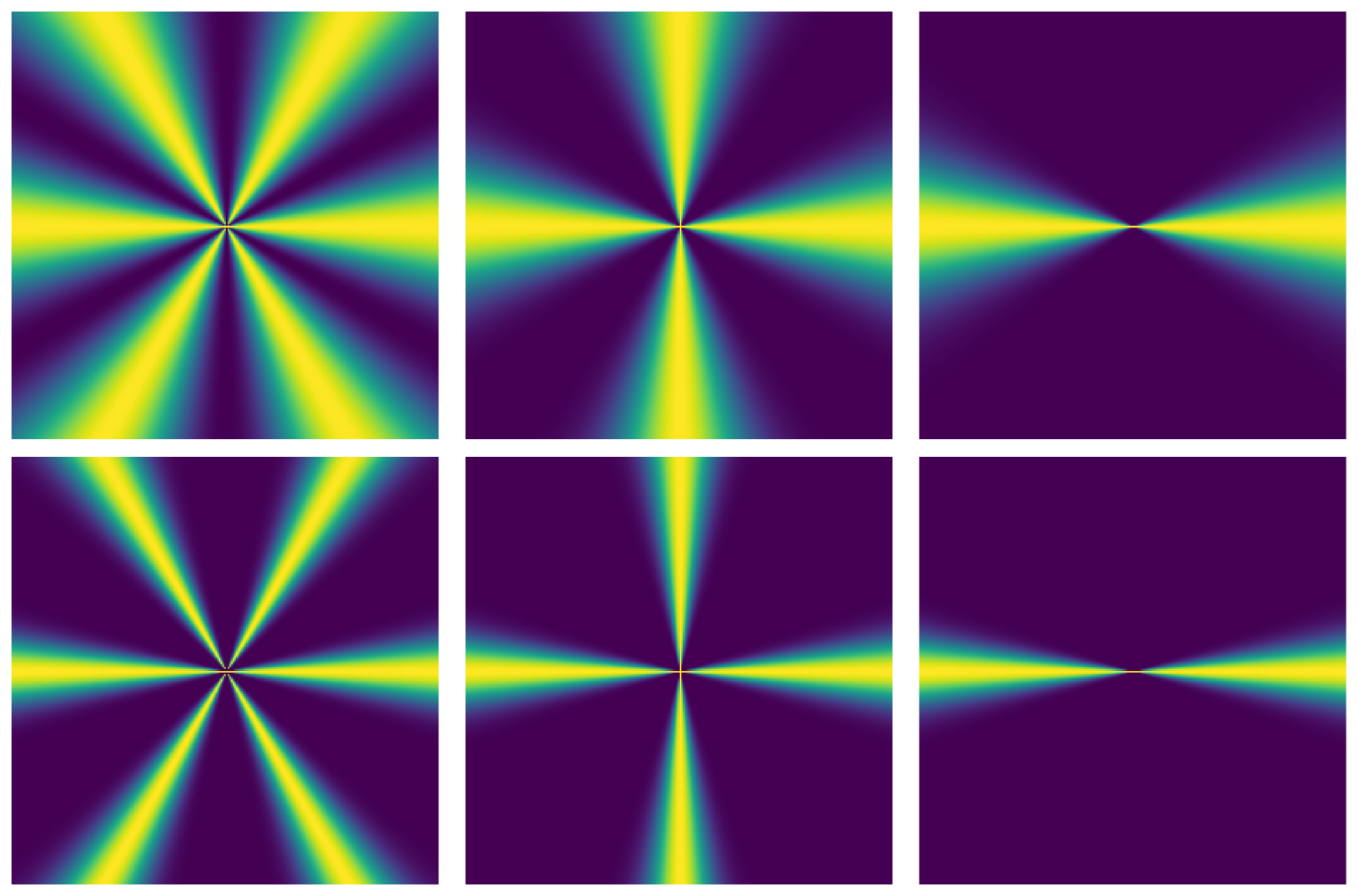}
    \caption{Comparison of the SDF for different orders of rotational symmetry and peak widths.
    From left to right, the columns correspond to $\nfolds=6,4,2$. From top to bottom, the rows are $\fwhm=20^\circ$ and $\fwhm=10^\circ$.
    }
    \label{fig:compare-ks-and-ns}
\end{figure}

To estimate a good value for $k$ we can use the relation,
\begin{equation}
\label{eq:fwhm-eq}
k = -\frac{\ln{2}}
{\ln\left[
\cos\left(
\frac{\nfolds \cdot \fwhm}{4}
\right)
\right]},
\end{equation}
where \fwhm is the full width at half maximum of the features of interest in the image.
Examples of the SDF for different orders and~$\fwhm$ values are shown in \fig{compare-ks-and-ns}.

\subsection{Creating the Symmetry-Adapted Filter (SAF)}
\label{sec:symmetry-adapted-filter}

The SAF, \SAF, is constructed from the SDF using the following procedure. The SkL example introduced in \sect{motivation} is used for illustration.

First, the grid on which the SDF is evaluated is created and taken to be the same as that of~\intensity.
Next, the symmetry center of \intensity is identified.
In the SkL example, the symmetry center is located approximately at $x_c = 64$ and $y_c = 62$ and is shown as a red dot in \fig{algorithm-overview}(a).
Then the SDF is calculated on the defined grid, centered on the symmetry center.
A mask  is then applied.
The mask is designed to suppress signals that are not relevant to the analysis.
In the SkL example, the signal of interest lies within a fixed radial annulus about the symmetry center. Therefore, an annular mask with fixed inner and outer radii is used to isolate the Bragg peaks while suppressing signal transmitted through the beam stop.
The mask is shown in \fig{algorithm-overview}(c), and the resulting SAF is shown overlaid on a SANS pattern in \fig{algorithm-overview}(d).

\subsection{Overlap Score}
\label{sec:overlap}

We define a generalized overlap score by taking the inner product of the SAF with the image data $\intensity$,
\begin{equation}
\label{eq:overlap-general}
\overlapeq.
\end{equation}

\subsection{Extracting Orientation Angle}
\label{sec:extracting-orientation-angle}

The overlap score is maximized when the filter aligns with bright features in the image with the right symmetry.
We can therefore use it to find the orientation angle of such features of interest, for example, the Bragg peaks in the~\skl study, by  evaluating
\begin{equation}
    \varphi_0^* = \operatorname*{argmax}\limits_{\varphi_0} \mathcal{O}(\varphi_0).
\end{equation}

Any approach that maximizes \overlap can be used.  In our \skl example we used a brute-force search. The orientation angle of the \nfolds-fold symmetric data is found by computing \overlap for \vphi values in the range $[0,2\pi/\nfolds)$ in user-specified step-sizes (usually one degree or so).  \phimax can also be found using gradient descent, which is used in~\sect{tem}.

\subsection{Orientation Extraction for Multiple Overlapping Symmetric Features}
\label{sec:multiple-signals}

\begin{figure}
    \centering
    \includegraphics[width=\linewidth]{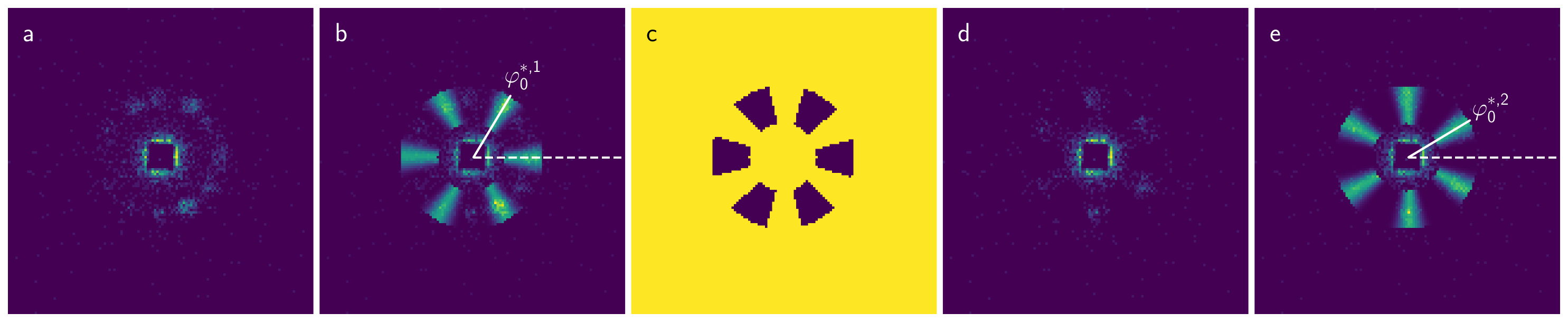}
    \caption{
    \textbf{(a)} A representative SANS pattern containing two symmetric signals with different orientation angles from the \skl experiment.
    \textbf{(b)} \Sof{\phimaxone} overlaid on the SANS pattern to highlight the stronger signal.
    \textbf{(c)} The mask created by thresholding \Sof{\phimaxone} at~0.03 via \eq{maskfilter}.
    \textbf{(d)} The SANS pattern with the mask in (c) applied, showing that the dominant signal is masked out but the weaker signal remains.
    \textbf{(e)} The masked SANS pattern in (d) overlaid with \Sof{\phimaxtwo} to highlight the weaker signal.
    }
    \label{fig:multiple-signals}
\end{figure}

Some images, such as the SANS pattern in \fig{multiple-signals}(a), contain multiple symmetric signals that share a common symmetry center and radial distance from the center but have different orientation angles.
This originates from multiple \skl domains in this case.

We would like a strategy for finding the angles of the majority and minority phases.
To extract the orientation angle of each feature, the algorithm is applied iteratively.

First, the orientation angle of the strongest signal, \phimaxone, is determined by maximizing the overlap score as described above.
\fig{multiple-signals}(b) shows $\Sof{\phimaxone}$ overlaid on a SANS pattern with multiple symmetric signals.

Once \phimaxone\ is found, the dominant signal is suppressed to allow extraction of the next strongest feature, \phimaxtwo.
To suppress the first signal, a binary mask, which we call the majority-phase mask \mapmask, is created by thresholding $\Sof{\phimaxone}$. A threshold value, \threshold, is defined and the mask is set to zero when $\Sof{\phimaxone}>\threshold$,
\begin{equation}
\label{eq:maskfilter}
\mapmask =
\begin{cases}
0 & \text{if } \Sof{\phimaxone} > \threshold \\
1 & \text{otherwise.}
\end{cases}
\end{equation}
The majority-phase mask used for the image in
\fig{multiple-signals}(a) for $\threshold=0.03$ is shown in \fig{multiple-signals}(c).
The second orientation angle, \phimaxtwo, is then determined by maximizing the minority-phase overlap score, \overlapsub{mip}, that now incorporates the majority-phase mask, \mapmask,
\begin{equation}
    \overlapsub{mip} = \SAF \cdot [\mapmask \odot \intensity],
\end{equation}
where $\odot$ indicates the Hadamard product and $\cdot$ the inner product.
The resulting image after applying \mapmask to the \skl SANS pattern is shown in  \fig{multiple-signals}(d) and
\Sof{\phimaxtwo} overlaid on the SANS pattern in \fig{multiple-signals}(e).

In principle, this iterative procedure can be extended to any number of symmetric features with the same \nfolds symmetry. In the datasets analyzed here, at most two \nfolds-symmetric features were present.

\subsection{Signal Detection Using the Second Derivative}
\label{sec:peak-detection}

\begin{figure}
    \centering
    \includegraphics[width=\linewidth]
    {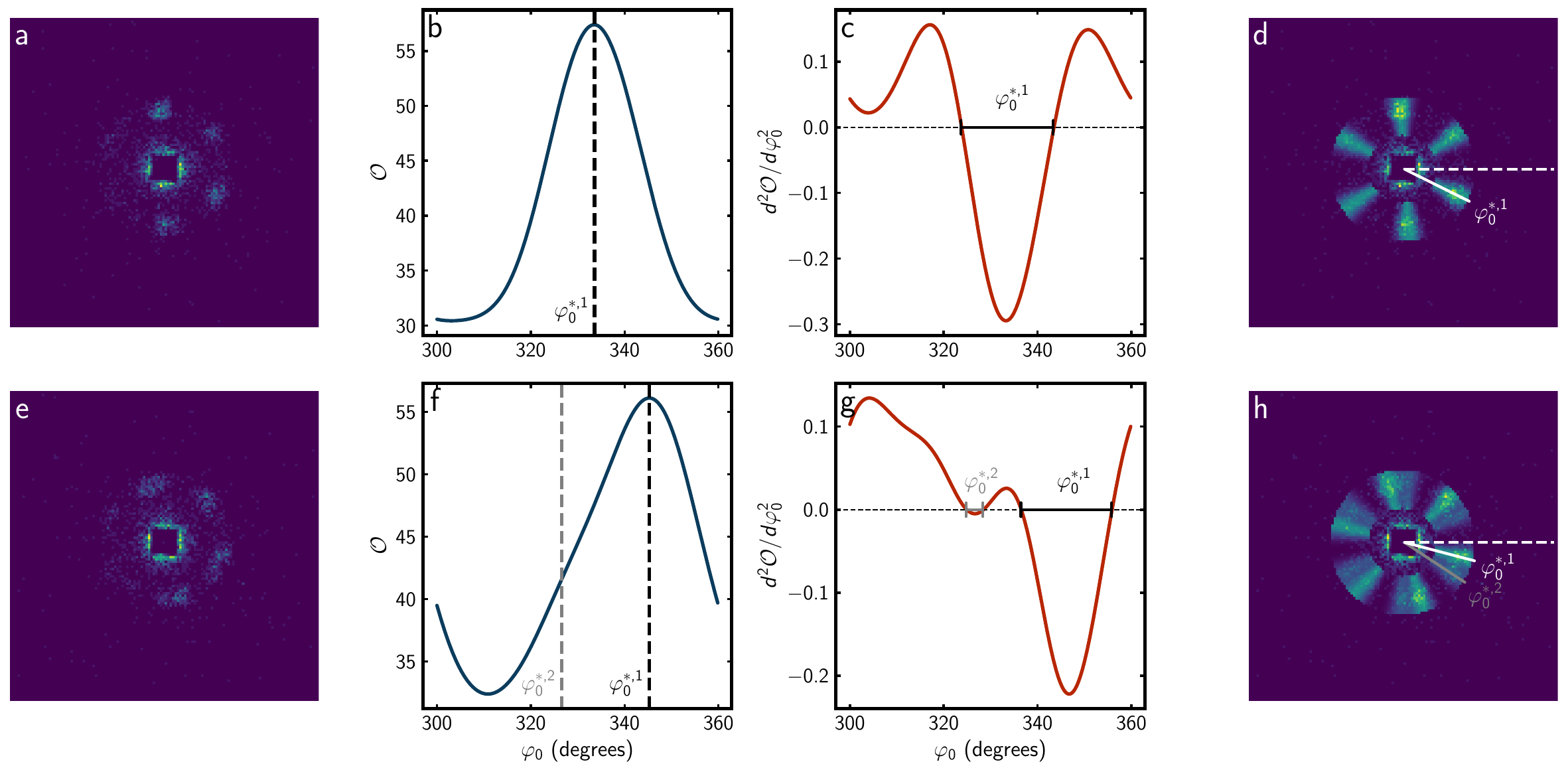}
    \caption{
    \textbf{(a)} A SANS pattern from a \skl containing one sixfold symmetric signal.
    \textbf{(b)} The overlap score calculated from the data in (a). The black vertical dashed line corresponds to the orientation angle of the single sixfold signal, \phimaxone.
    \textbf{(c)} The second derivative of the overlap score in (b). Two x-intercepts signify the presence of one sixfold signal. The orientation of that signal is located between those x-intercepts.
    \textbf{(d)} The SANS pattern in (a) overlaid with \Sof{\phimaxone}.
    \textbf{(e)} A SANS pattern from the same \skl sample under different conditions, containing two unresolved sixfold symmetric signals.
    \textbf{(f)} The overlap score calculated from the data in (e). The black vertical dashed line corresponds to the orientation angle of the stronger signal, \phimaxone. The gray dashed line corresponds to the orientation angle of the weaker signal, \phimaxtwo, that was found after the full analysis. Here, the weaker signal produces a slight shoulder in the overlap score.
    \textbf{(g)} The second derivative of the overlap score in (f). Here, four x-intercepts are found indicating the presence of two sixfold signals. The orientation angle for each signal is located between the respective x-intercepts.
    \textbf{(h)} The SANS pattern in (e) overlaid with \Sof{\phimaxone} and \Sof{\phimaxtwo}.
    }
    \label{fig:second-der}
\end{figure}

When multiple symmetric signals are present, the peaks in the image sometimes overlap and become difficult to resolve.
This is illustrated in \fig{second-der} where diffraction patterns with a single signal, (a), and two unresolved signals, (e), can be compared.
The images look highly similar.
In the single signal case, the overlap score exhibits a single distinct maximum (\fig{second-der}(b)).
In the unresolved case there is also a single maximum in \overlap, \fig{second-der}(f), and the weaker signal does not generate a second maximum.
Instead, its contribution appears only as a shoulder on the left of the dominant peak.
In cases like this, it is not possible to identify the presence of the unresolved signal from the maxima in \overlap alone.

We seek a way of automatically identifying when multiple overlapped signals are present in order to extract their orientation angles.
Because \SDF is smooth and the overlap score is computed from inner products over the \vphi-range, \overlap is also smooth, allowing numerical derivatives to be computed with minimal sensitivity to noise in \intensity.
We therefore use the second derivative, \ddoverlap, to detect changes in curvature introduced by additional signals.

When a second signal is present, an additional region of negative curvature is introduced in \overlap.
The new region of negative curvature introduces inflection points in \overlap, which are represented as x-intercepts in \ddoverlap.
Therefore, each distinct signal in the data will introduce two x-intercepts in \ddoverlap even if the peak itself is not resolved as in \fig{second-der}(f).

The second derivative, \ddoverlap, is illustrated in \fig{second-der}(c) and (g).
The sample with two unresolved domains has four x-intercepts in \ddoverlap  (\fig{second-der}(g)) where the sample with a single domain shows only two x-intercepts (\fig{second-der}(c)).
The orientation angle of each signal lies between the corresponding pair of inflection points.

By counting the number of zero-crossings in \ddoverlap, we can automatically identify images containing multiple symmetric signals.
Once identified, the iterative extraction procedure described in \sect{multiple-signals} can be used to recover the orientation angle of each signal.
If a gradient-based local optimization method is used, the negative minimum in \ddoverlap between each pair of zero-crossings, which lies roughly at their midpoint, provides a suitable initial guess for the peak angle in the nonlinear least-squares optimization.
This approach proved effective for datasets containing multiple symmetric signals, such as those reported in~\cite{chalus2026therbino}.

\section{Example Applications}

In this section, we go through two example cases where the SAF algorithm was used.
\sect{skl} covers the application that motivated the SAF algorithm and has been discussed throughout this paper, and \sect{tem} applies the SAF algorithm to an atomically resolved transmission electron microscopy (TEM) image.

\subsection{Skyrmion Lattice}
\label{sec:skl}

\begin{figure}
    \centering
    \includegraphics[width=\linewidth]{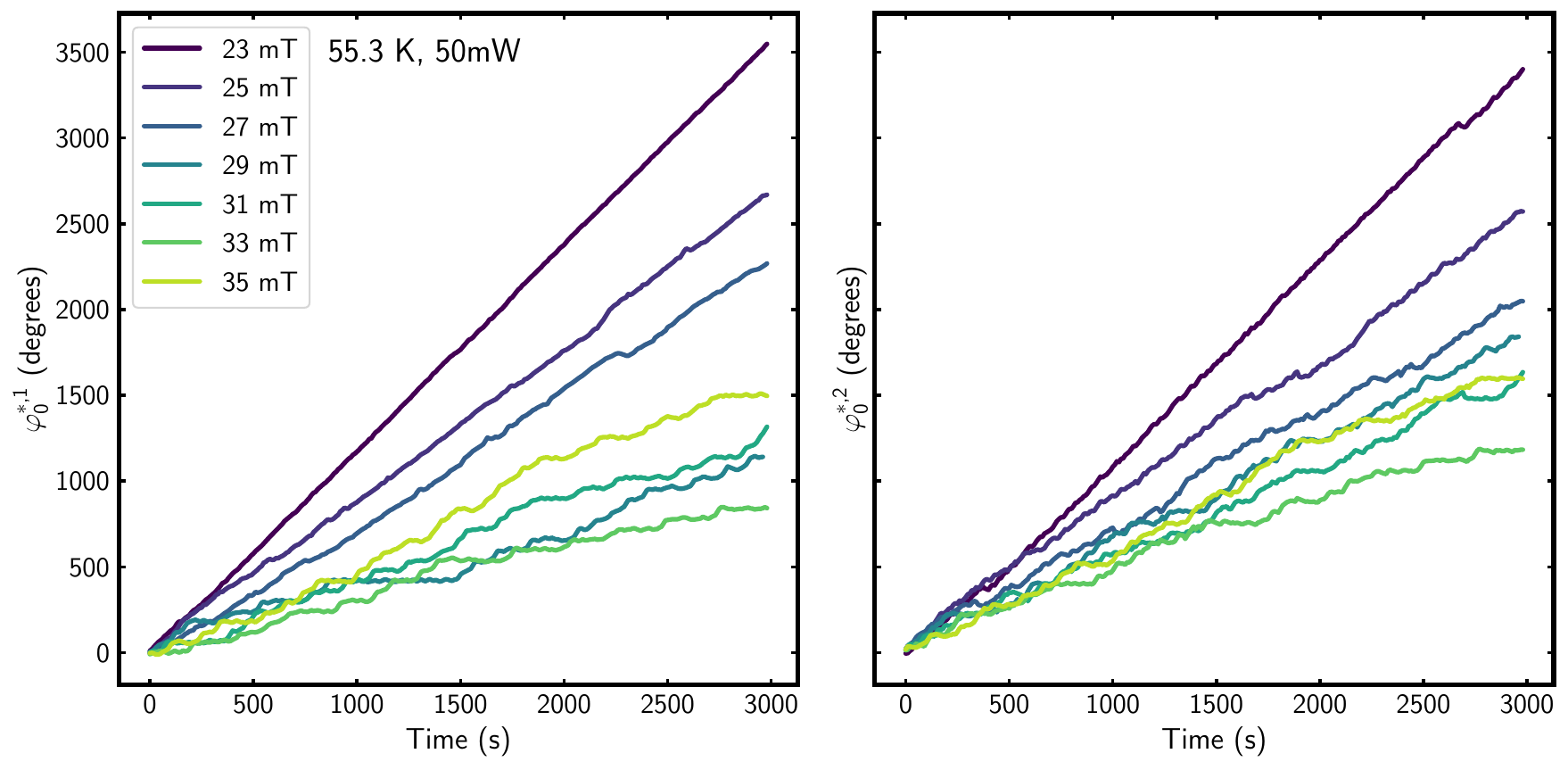}
    \caption{The orientation angles of both sixfold signals versus time of a representative magnetic field sweep measured at 55.3~K and an applied heating power of 50~mW.}
    \label{fig:offset-plots}
\end{figure}

The SAF algorithm was developed for and applied to the SkL data discussed above in Sections~\ref{sec:motivation}-\ref{sec:algorithm-details}.  Here we explain the science that was learned from application of the SAF.  More information can be found in \cite{chalus2026therbino}.
The algorithm allowed us to automatically analyze 69~movies, totaling 20,000+ time-series diffraction patterns, many of which contain two sixfold signals, where one was often not bright or well-resolved. An example of this is shown in \fig{second-der}(e).
Depending on the experimental conditions, the movies either showed continuous linear rotation or a ``stick-slip'' rotation, in which the sixfold signal would rotate intermittently, alternating between periods of rotation and stasis.
We therefore sought to quantitatively characterize this behavior by extracting the orientation angles versus time of both signals to better understand the underlying physics of SkL dynamics under a driving field.
The large number of images, poor peak resolution, and superposition of the two sixfold signals made manual analysis impractical.
To automate this analysis, the SAF algorithm was used to extract the orientation angles of both signals in all the images throughout the movies, capturing both linear and stick-slip rotation depending on the conditions.
The resulting curves from a representative magnetic field sweep dataset are shown in \fig{offset-plots}.
From these curves, the crossover from  stick-slip to linear rotation behavior can clearly be seen as the field strength decreases and the angular velocity can be measured as the slope of the curve.
Additionally, correlations between the plateau regions of \phimaxone and \phimaxtwo were observed.
The quantitative results obtained using the SAF approach gave critical insights into the  underlying physics that affects the SkL~\cite{chalus2026therbino}.
Movies of this dataset can be found at \url{https://tinyurl.com/mt5wjukj}.

\subsection{Extracting Atomic Neighborhood Orientation from Atomically Resolved Images}
\label{sec:tem}

Having developed the SAF algorithm for the SkL case discussed above, we show how the algorithm can be generalized to any type of image analysis involving the extraction of orientational information.
For example, we used the SAF algorithm here to find orientations of nearest neighbor arrangements of atoms in atomically resolved transmission electron microscopy (TEM) images.
Doing this enabled the extraction of various kinds of information from the image, such as long-range strain, automatic identification of  grains and lattice defects, and visualization of low-angle grain boundaries.
In this application, the SAF algorithm is applied many times to the same image, centered iteratively on each atom in the image.
This process returns the local orientation angle referenced to each atom.
The SAF algorithm was applied to a simulated atomically resolved TEM image from~\cite{berkelsIdentificationGrainBoundary2007} and a real TEM image from~\cite{matcaTEM}.
The simulated image is reproduced in \fig{tem-mask}(a) and will be used to demonstrate the application of the algorithm to getting the orientation of local atomic environments.
The image from~\cite{matcaTEM} is discussed in \sect{real-tem}.

\subsubsection{Finding the Atomic Coordinates}
\label{sec:atom-centers}

The goal is to find the orientation of the nearest neighbor shell of atoms around each atom.
Therefore, the first step is to find the coordinates of every atom in the image.
The coordinates will be used to iteratively center the SAF.
A rough location was determined using the Python package \atomap~\cite{atomap} that uses a center-of-mass approach to determine approximate atomic coordinates.
These coordinates provide an initial estimate of the position of each atom.
A 2D Gaussian is then fit to each atom using this initial estimate as the starting position.
The centers of the fitted Gaussians provide the refined atomic coordinates.
The refined atomic coordinates plotted on top of the TEM image are shown in \fig{tem-mask}(a).

With all the atomic coordinates found, the SAF algorithm is applied by centering it on each atomic site.

\subsubsection{Determining the Mask}

Following the algorithm described in~\sect{algorithm-details}, the next step is to design a mask that isolates the signal of interest.
Here, we aim to extract the orientation angle of the nearest-neighbor atomic arrangement surrounding a central atom.
Therefore, the signal of interest consists of the nearest-neighbor atoms surrounding the central atom, which we refer to as the atomic neighborhood.

\begin{figure}
    \centering
    \includegraphics[width=\linewidth]{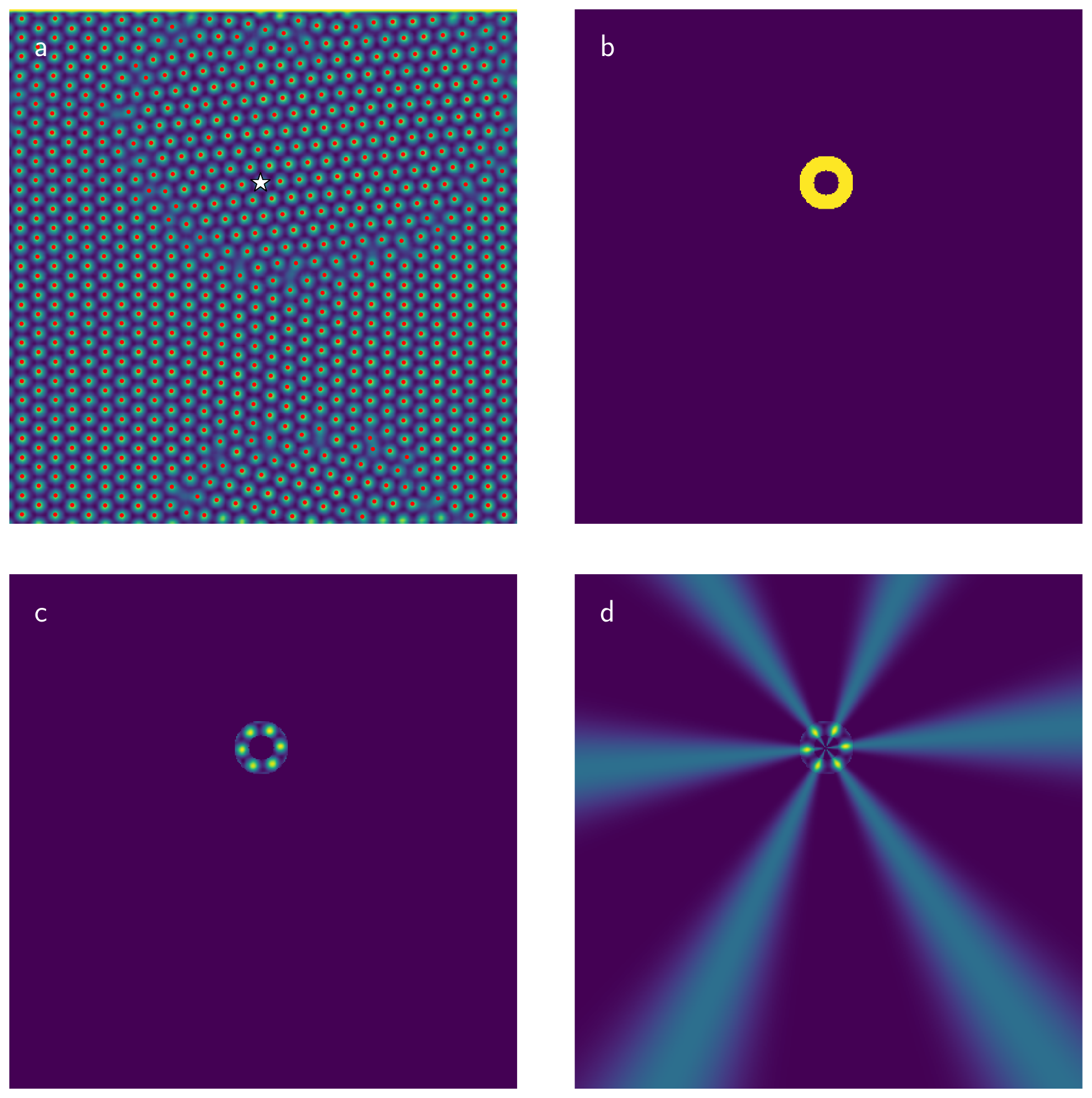}
    \caption{
    \textbf{(a)} The simulated TEM image from \cite{berkelsIdentificationGrainBoundary2007}. The red dots indicate the atomic centers found by \atomap.
    The white star marks an arbitrarily chosen atom to illustrate the application of the SAF to a single site.
    \textbf{(b)} The annular mask appropriate for isolating the centered atom's atomic neighborhood.
    \textbf{(c)} The mask applied to the TEM image. Only signal from the atomic neighborhood remains.
    \textbf{(d)} \Sof{\phimax} overlaid on the masked TEM image. Here, the mask is applied to the image instead of the SDF for visualization purposes.
    }
    \label{fig:tem-mask}
\end{figure}

To define the mask, we choose an annular mask with an inner and outer diameter that captures the majority of the atomic intensity in the nearest neighbor shell when it is centered on an atomic site.
The mask used is shown in \fig{tem-mask}(b).
The image resulting from centering the mask on the atom marked with the white star in \fig{tem-mask}(a) is shown in \fig{tem-mask}(c).

\subsubsection{Determining the Orientation Angle}

Now, the orientation angle of the atomic neighborhood is determined as described in~\sect{algorithm-details}.
The optimized SAF overlaid on the masked TEM image is shown in \fig{tem-mask}(d).
Because the variation in orientation angle between atomic neighborhoods is small, a gradient descent approach is used to determine \phimax with starting values taken from neighboring solutions.

\subsubsection{Orientational Strain and Grain Mapping}
\label{sec:strain-mapping}

\begin{figure}
    \centering
    \includegraphics[width=\linewidth]{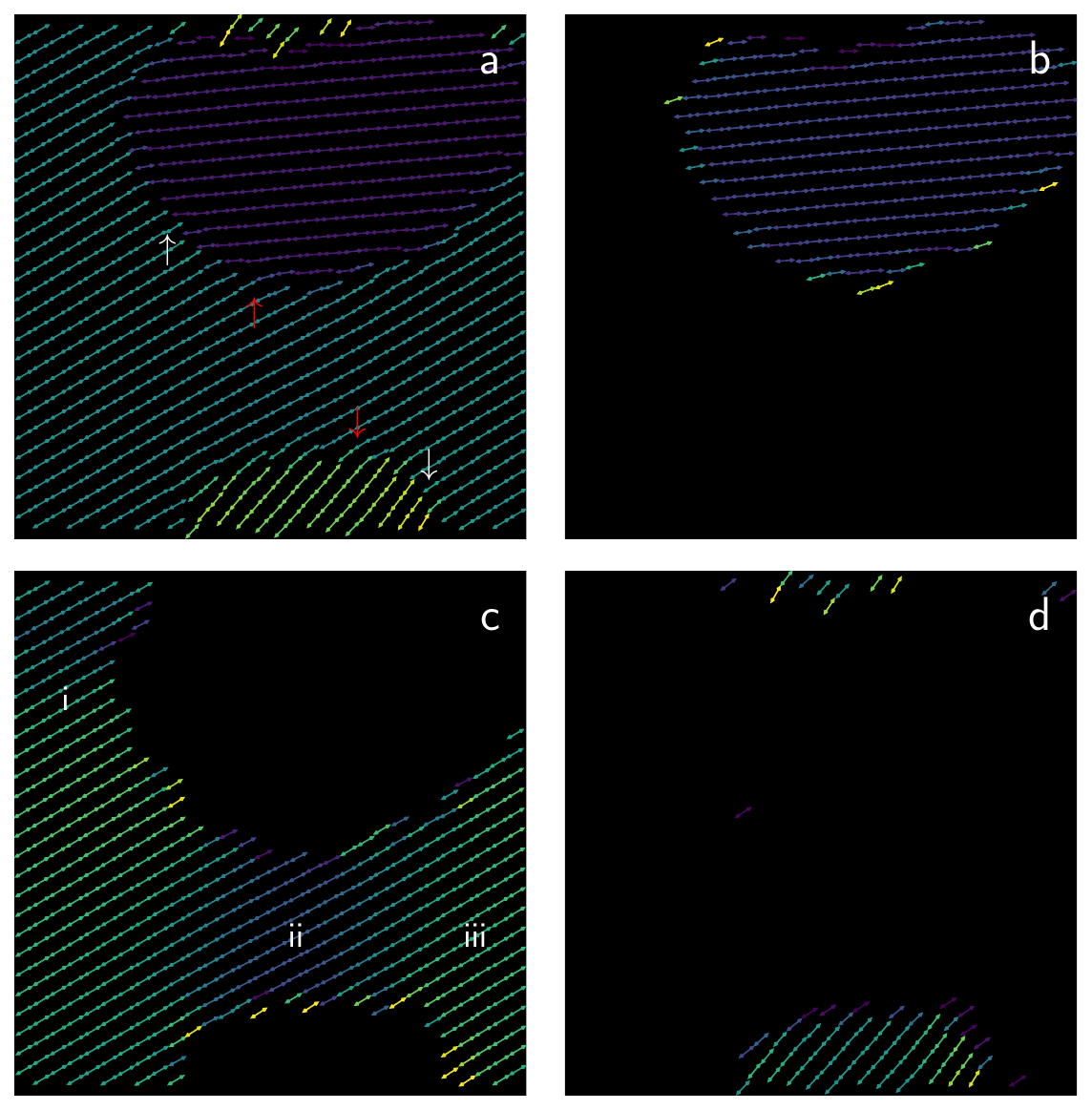}
    \caption{
    \textbf{(a)} The local neighborhood orientation map of the TEM image in \fig{tem-mask}(a). The center points of the arrows sit on the atomic sites. Each arrow's direction and color represents the orientation angle of the atomic neighborhood at that atomic site.
    \textbf{(b)--(d)} The atomic neighborhood orientation map in panel (a) segmented into three maps. The full angular range is split across the three panels to make small angular differences between grains distinguishable.
    }
    \label{fig:quiverplot}
\end{figure}

The process described above is then repeated across every atomic coordinate using the same SAF.
This results in each atomic coordinate having an associated angle corresponding to its atomic neighborhood orientation.
Because of the sixfold symmetry of the SAF, the extracted orientation angles are reduced modulo $60^\circ$, restricting the reported angles to the range $0^\circ$ to $60^\circ$.
The orientation angles can be visualized as quiver plots shown in \fig{quiverplot}.
The position of the center point of each arrow corresponds to the atomic coordinate where the SAF was centered.
The arrow direction and color encode the orientation angle of the atomic neighborhood.
This representation maps atomic neighborhood orientations across space, allowing variations associated with strain, defects, and grain boundaries to be directly observed.

In \fig{quiverplot}(a), the three grains in the TEM image show up as three distinct regions, due to the orientation of each grain being different.
This allows for domains with similar orientation angles to be easily identified.

\begin{figure}
    \centering
    \includegraphics[width=\linewidth]{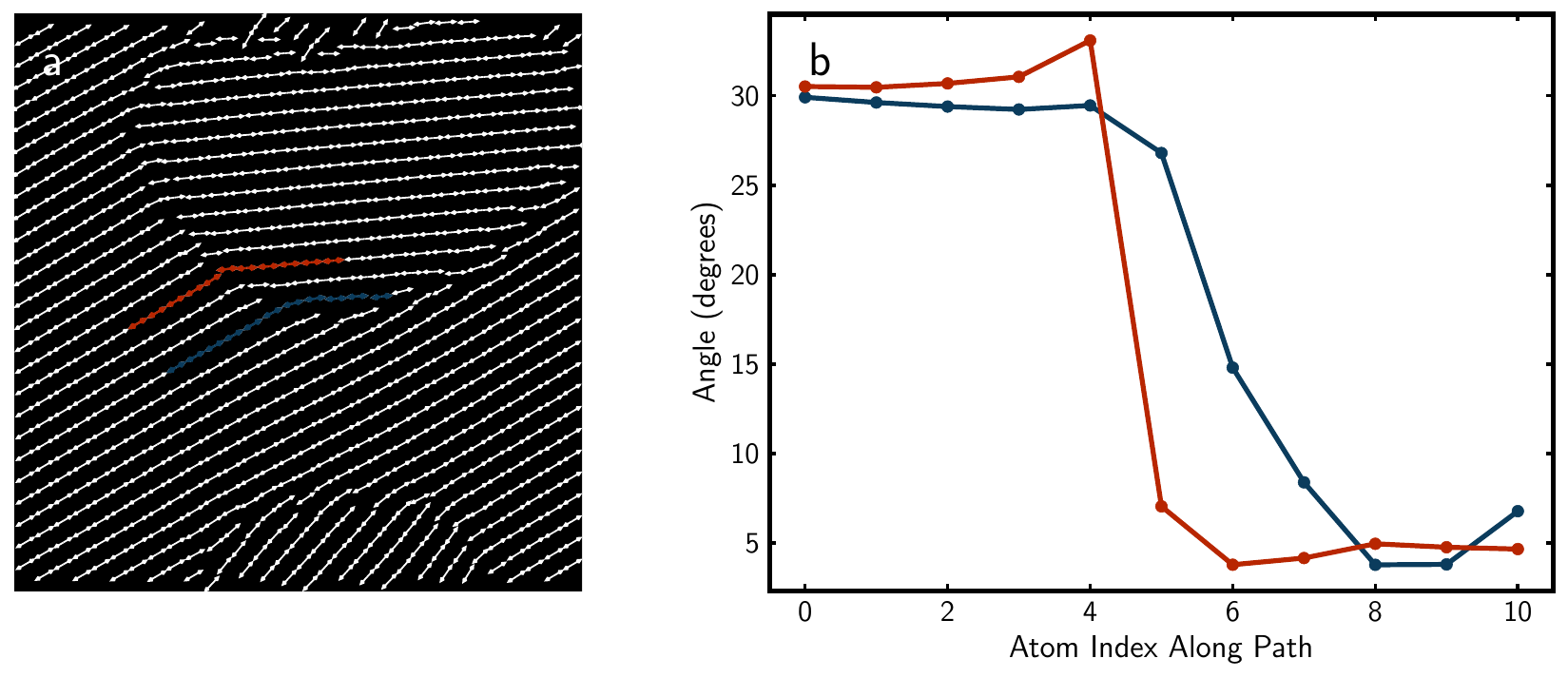}
    \caption{
    \textbf{(a)} The atomic neighborhood orientation map of the simulated TEM image. The colored arrows highlight where an atomic plane intersects a grain boundary at a low angle (blue) and high angle (red).
    \textbf{(b)} The orientation angle of atomic neighborhoods of the highlighted paths in (a) as a function of atom index along the path. The blue curve shows a gradual transition as the path crosses the grain boundary whereas the red curve is abrupt.
    }
    \label{fig:tem-cont-vs-abrupt}
\end{figure}

In addition to segmenting the image by grains, lattice orientational mismatch can be quantified as a function of position, for example, across a grain boundary.
\fig{quiverplot}(a) shows regions of abrupt mismatch between grains (white arrows) as well as regions of gradual orientation change (red arrows).
This orientational mismatch can be quantified directly as a change in angle across the grain boundary.
\fig{tem-cont-vs-abrupt} shows this angle change for an abrupt change (red) and a gradual change (blue).

In \fig{tem-cont-vs-abrupt}(a), we show the two different atom trajectories and in
\fig{tem-cont-vs-abrupt}(b) the angle change vs. atom index is plotted. This illustrates the kind of quantitative information that can be extracted from the neighborhood angle data.

Another feature highlighted by the quiver plots is the presence of subtle lattice distortions and strain.
These can be qualitatively identified from the spatial distribution of orientation angles, where color variations within a grain indicate local changes in lattice orientation.
This is illustrated in \fig{quiverplot}(b)--(d), with the clearest example shown in \fig{quiverplot}(c).
Here, subtle lattice distortions are visible as gradual changes in color, progressing from the top left (marked ``i''), through the center (marked ``ii''), to the bottom right (marked ``iii'').

\begin{figure}
    \centering
    \includegraphics[width=\linewidth]{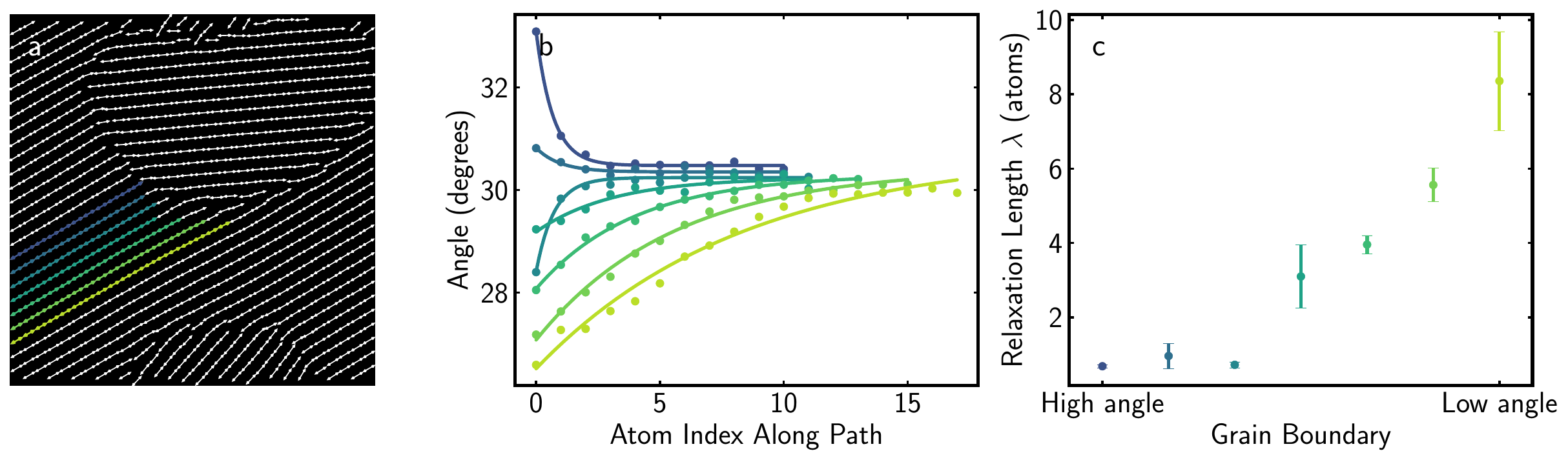}
    \caption{
    \textbf{(a)} The atomic neighborhood orientation map of the TEM image. The colored arrows highlight planes of atoms along a path that intersects the lower grain.
    \textbf{(b)} The orientation angle of atomic neighborhoods calculated from the SAF algorithm of the highlighted paths in (a) as a function of atoms along the path (dots). The lines are exponential fits of the form $\phimax(d) = \varphi_{0,\infty}^* + A \exp\!\left(-\frac{d}{\lambda}\right)$ where $\lambda$ represents the relaxation length in units of atoms.
    \textbf{(c)} The relaxation lengths plotted as a function of the paths traced in (a). The colors correspond to the respective paths.
    }
    \label{fig:tem-relaxation-length}
\end{figure}

As atomic planes extend away from a grain boundary and into the bulk, the lattice gradually relaxes toward its unstrained equilibrium configuration.
The local orientation angles can be used to quantify this relaxation by fitting an exponential decay, from which a characteristic ``relaxation length'' is extracted.
The relaxation length describes the distance over which an atomic plane approaches its equilibrium orientation.
The extraction of relaxation lengths for several atomic planes is shown in \fig{tem-relaxation-length}.

The selected atomic planes are highlighted in different colors in \fig{tem-relaxation-length}(a).
These planes were chosen to span a range of intersection angles with the grain boundary, from high to low, while extending sufficiently far into the bulk to reach an unstrained equilibrium state.
\fig{tem-relaxation-length}(b) shows the local orientation angles along each atomic plane, plotted as individual points.
The data are fit with an exponential decay of the form $\phimax(d) = \varphi_{0,\infty}^* + A \exp\!\left(-\frac{d}{\lambda}\right)$, where $\lambda$ represents the characteristic relaxation length.
The extracted relaxation lengths are plotted as a function of the atomic plane intersection angle in \fig{tem-relaxation-length}(c), progressing from high to low angles.
The observed increase in relaxation length at lower intersection angles may arise from a tendency of these planes to align with the intersecting grain, resulting in strain that extends farther into the bulk than for planes with higher intersection angles.

\subsubsection{Application to an Experimental TEM Image}
\label{sec:real-tem}

\begin{figure}
    \centering
    \includegraphics[width=\linewidth]{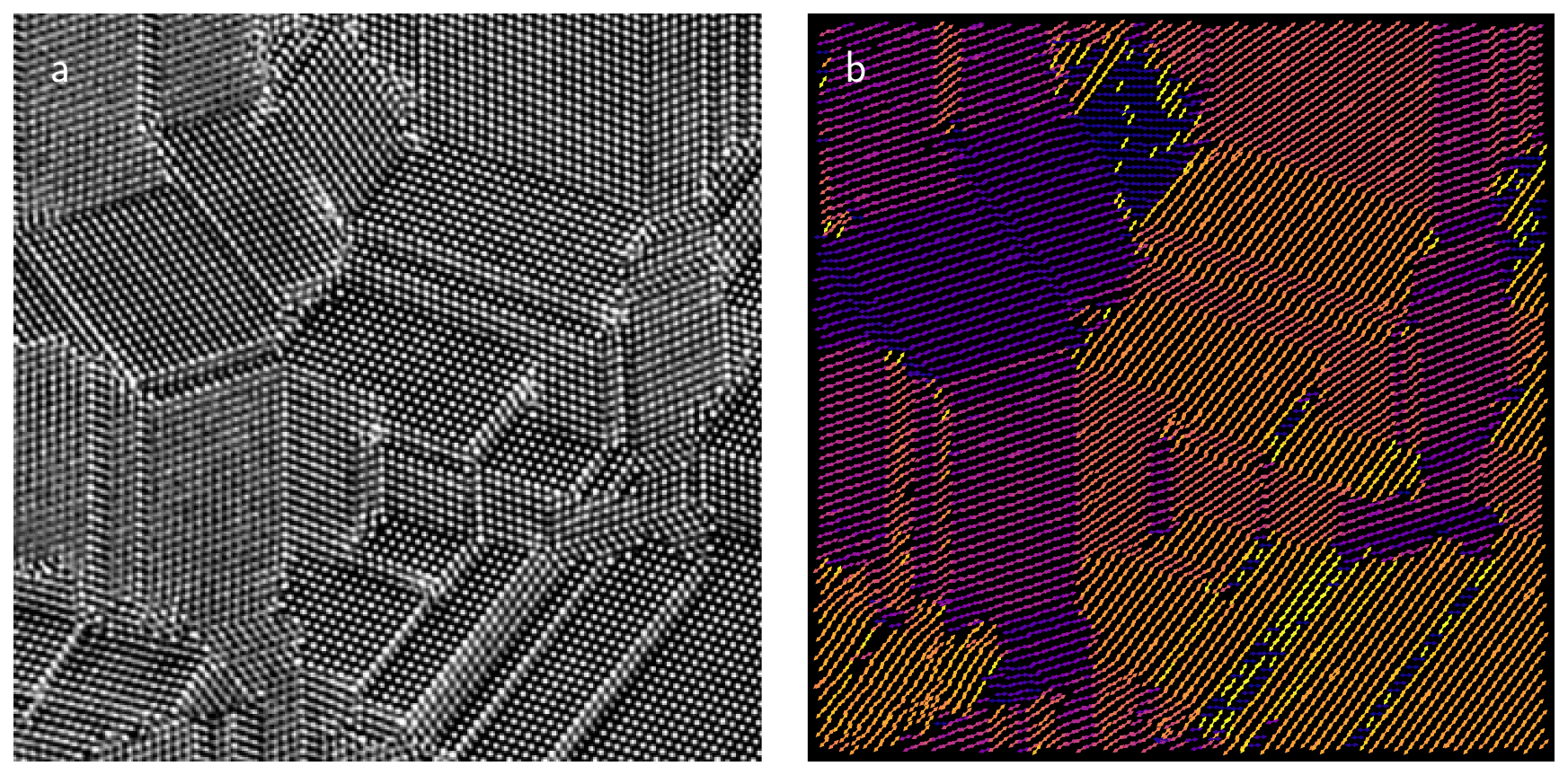}
    \caption{
    \textbf{(a)} A TEM image of polycrystalline silicon adapted from~\cite{matcaTEM}, used with permission from MATCA.
    \textbf{(b)} The atomic neighborhood orientation map of (a).}
    \label{fig:real-tem}
\end{figure}

Here, we apply the same approach to an experimental TEM image of polycrystalline silicon.
The image, taken from~\cite{matcaTEM}, is shown in \fig{real-tem}(a).
The corresponding atomic neighborhood orientation map is shown in \fig{real-tem}(b).
Compared with the simulated image analyzed above, this dataset exhibits several notable differences.
First, it contains a larger number of grains.
Second, the grains exhibit extensive twinning.
Finally, there are variations in atomic resolution across the image, with some grains exhibiting more clearly resolved atomic positions than others.
By applying the SAF algorithm, we can segment the image into grains with similar lattice orientations and identify twinning from the spatial distribution of local orientation angles.

\begin{figure}
    \centering
    \includegraphics[width=\linewidth]{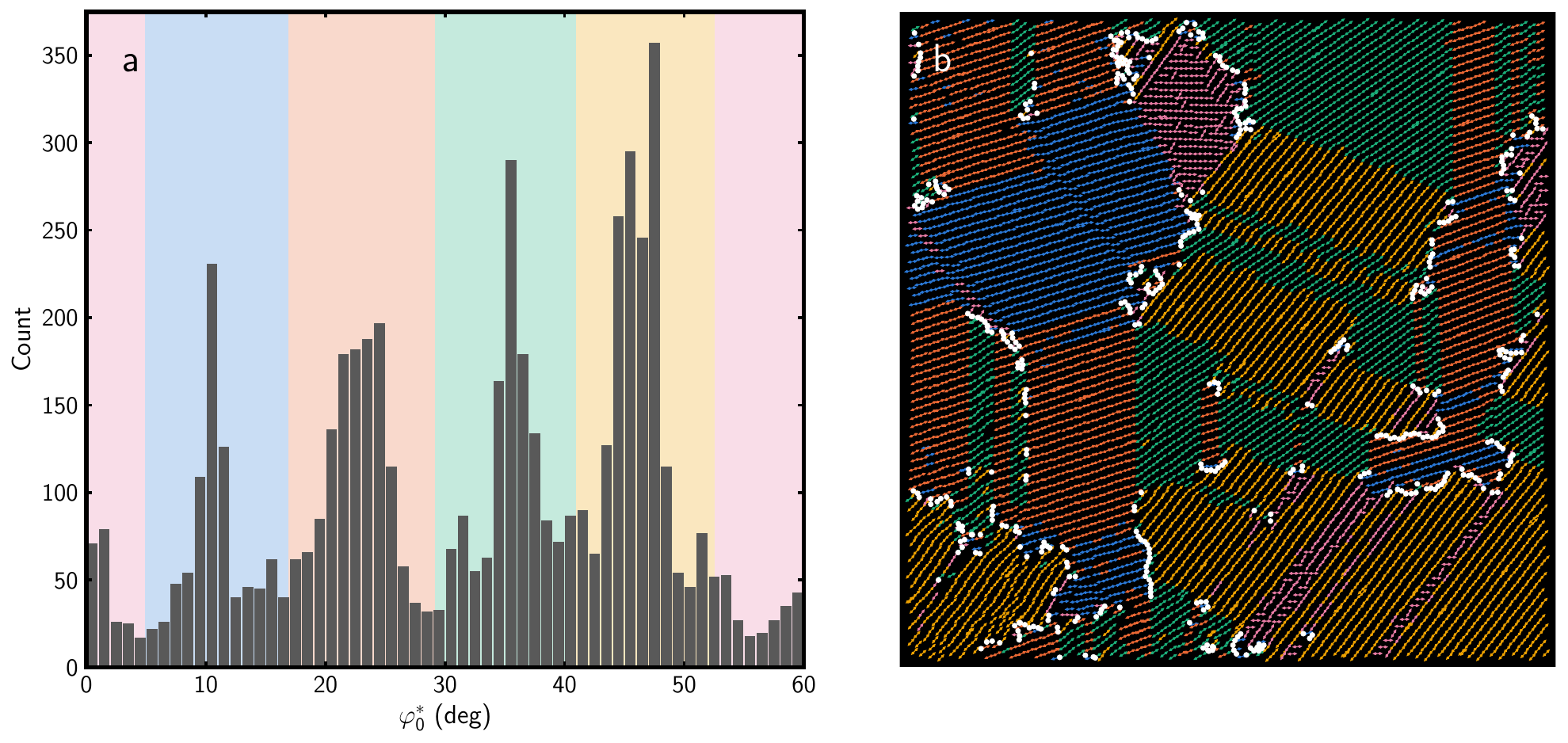}
    \caption{
    \textbf{(a)} Histogram of atomic neighborhood orientation angles, showing five distinct modes identified using $k$-means clustering.
    Shaded regions indicate the five clusters.
    \textbf{(b)} Spatial orientation map, with orientation vectors colored according to their assigned $k$-means cluster in (a).
    White markers indicate regions where grains do not border grains from neighboring orientation clusters.}
    \label{fig:real-map}
\end{figure}

To identify grains with similar lattice orientations, we first examine the distribution of local orientation angles across the image.
A histogram of these angles is shown in \fig{real-map}(a).
The histogram exhibits five distinct modes, with their centers separated by approximately $15^\circ$.
Given the sixfold symmetry of the lattice, this angular separation is characteristic of twinning and indicates a high density of twin boundaries.
To spatially identify regions belonging to each orientation mode, we apply a $k$-means clustering algorithm with the number of clusters set to five~\cite{pedregosaScikitLearn2011}.
This choice is motivated by the five distinct modes observed in the orientation histogram, which represent the dominant lattice orientations across the image.
In general, for segmenting an image by grain orientations, the number of clusters should equal the number of grain orientations in the image.
The resulting clusters are indicated by different colors in \fig{real-map}(a).
The corresponding spatial distribution is shown in \fig{real-map}(b), where each orientation vector is colored according to its assigned cluster.
\fig{real-map}(b) shows that many grains border grains belonging to neighboring orientation clusters.
For example, orange grains frequently border blue and green grains, while yellow grains border pink and green grains.
In regions where this pattern is not observed, such as the lower-left corner of the image, increased disorder and lattice mismatch are apparent.
The SAF orientation map makes these regions easy to identify; they are highlighted by white markers in \fig{real-map}(b).

\section{Future Directions}
\label{sec:future-directions}

Because the SAF algorithm only requires a symmetry center and the order of the symmetry, its application can extend to other types of data where symmetric features are present.
For example, it could be applied to scanning small- and wide-angle X-ray scattering (SAXS/WAXS) experiments to map feature orientation as a function of sample position.
In these experiments, every scan position produces its own scattering pattern, and aligned structures, such as the collagen fibrils in bone, give twofold-symmetric arcs whose angle is the local fibril orientation~\cite{liebiNanostructureSurveysMacroscopic2015}.
Where fibrils with different orientations overlap in the beam, the multiple-signal procedure described in \sect{multiple-signals} could separate them.

Electron diffraction datasets could also benefit from the SAF algorithm.
In four-dimensional scanning transmission electron microscopy (4D-STEM), a diffraction pattern is recorded at every probe position~\cite{ophusFourDimensionalScanning2019}.
Applying the SAF to zone-axis patterns would give fast in-plane orientation maps that complement existing orientation-mapping tools~\cite{rauchAutomatedCrystalOrientation2014, cautaertsFreeFlexibleFast2022}.
Similarly, in low-energy electron diffraction (LEED), the SAF could track the orientation of rotated surface domains during film growth~\cite{hattabGrowthTemperatureDependent2011}.

The SAF algorithm could also be applied to simulation data, such as molecular dynamics.
For example, simulations of vortex lattices with anisotropic interactions produce domains with different orientations, giving structure factors with multiple sixfold signals~\cite{olszewskiRotationalTransitionDomain2020, minogueActivatedVortexLattice2023}.
The SAF could track these orientations throughout the simulation, as was done for the SkL in \sect{skl}.

The SAF algorithm could also support existing model-based analysis.
Because it finds the number of symmetric signals and their orientation angles, its output could seed peak fits in SANS programs such as GRASP~\cite{dewhurstGraphicalReductionAnalysis2023}, setting how many peaks to fit and where to start without manual input.

Other filter shapes could extract other information from the data.
For example, rotating the SAF finds the angular position of the peaks, while replacing the fixed annular mask with an annulus with a Gaussian radial profile and scanning its radius would find their radial position.
Used together, the two would give full peak positions in a noise-tolerant way that uses the known symmetry of the data.

More broadly, any high-throughput experiment that produces images with orientation-dependent features could be reduced to a set of orientation angles and signal counts with little manual effort.

\section{Conclusion}

This work presents the symmetry-adapted filter (SAF) algorithm for extracting the orientation angle of features in 2D images.
The SAF encodes a known or hypothesized rotational symmetry of the image, is centered on the symmetry center, and is masked to the region of interest.
Rotating the SAF and maximizing its inner product with the image yields the orientation angle of features in the image with the target symmetry.
The method is deterministic and needs no training data, so it can be adapted to a new dataset by choosing the symmetry order, peak width, and mask.
When several signals with the same symmetry overlap, the second derivative of the overlap score detects them, and masking out the strongest signal allows each orientation angle to be extracted in turn.

We demonstrated the SAF algorithm on two very different types of data.
For SANS of a driven SkL in \ce{Cu2OSeO3}, it extracted the orientation angles of two coexisting sixfold signals from more than 20,000 patterns across 69 movies, many with weak or unresolved peaks.
This revealed the crossover from stick-slip to continuous rotation and gave the angular velocity of the lattice~\cite{chalus2026therbino}.
For atomically resolved TEM images, centering the SAF on each atom gave a map of the local neighborhood orientation that identifies grains, distinguishes abrupt from gradual grain boundaries, and quantifies how the lattice relaxes away from a boundary.
The same approach also worked on an experimental TEM image of polycrystalline silicon.
Installable code for the SAF algorithm can be found at \url{https://github.com/diffpy/diffpy.saf}.

Because the SAF algorithm only needs a symmetry center and the order of the symmetry, it should be useful for other high-throughput measurements with symmetric features, such as scanning SAXS/WAXS and electron diffraction, and for seeding model-based peak fits.
More generally, it offers a simple way to reduce large image datasets to orientation angles and signal counts with little manual effort.

\section{Author Contributions}
T.S. and S.J.L.B. proposed the initial idea for the symmetry-adapted filter design.
C.M. and Y.C. refined the algorithm and performed the data analysis.
N.C. collected the SANS data at the Paul Scherrer Institute and helped analyze the results of the algorithm.
M.R.E. supervised the neutron scattering work.
S.J.L.B. supervised the project.
C.M. and S.J.L.B. wrote the manuscript, and all authors contributed to and reviewed it.

\section{Acknowledgements}

Work in the Billinge Group was supported by U.S. Department of Energy, Office of Science, Office of Basic Energy Sciences (DOE-BES) under contract No. DE-SC0024141.
The Eskildsen Group acknowledges the U.S. Department of Energy, Office of Basic Energy Sciences, under Award No. DE-SC0005051 (N.C., M.R.E.) for the SkL data acquisition.
T.S. was supported by the Air Force Office of Scientific Research through the Young Investigator Program (AFOSR YIP), Award No.~FA9550-24-1-0067.
SANS data were collected at the SANS-I instrument at SINQ at the Paul Scherrer Institute.
We would like to thank the National Centre of Competence MATCA (www.matca.cz), coordinated by the Institute of Physics of the Czech Academy of Sciences, for permission to use the HRTEM image of polycrystalline silicon.

\bibliographystyle{iucr}
\bibliography{bg-pdf-standards,billinge-group-bib,cm_skyrmion_lattice}

\end{document}